\documentclass[11pt]{article}

\usepackage[margin=1in]{geometry}
\usepackage{times}
\usepackage[T1]{fontenc}
\usepackage[utf8]{inputenc}
\usepackage{graphicx}
\usepackage{booktabs}
\usepackage{amsmath,amssymb}
\usepackage{caption}
\usepackage{subcaption}
\usepackage{microtype}
\usepackage{natbib}
\usepackage{xurl}
\usepackage[colorlinks=true,linkcolor=blue,citecolor=blue,urlcolor=blue]{hyperref}
\usepackage{xcolor}
\usepackage{authblk}
\usepackage{enumitem}
\usepackage{multirow}
\usepackage{listings}
\title{\textbf{Diffs vs.\ Whole Files: An Empirical Comparison of Iterative\\
Edit-Based and Direct Generation for Flutter/Dart Code Models}}

\author[1]{Andrej Andrejev}
\affil[1]{Independent researcher (bbidpa) \\ \texttt{andrew.andrejev@gmail.com}}
\date{\today}

\begin{document}
\maketitle

\begingroup
\renewcommand{\thefootnote}{}
\footnotetext{This work is licensed under a Creative Commons Attribution 4.0
International License (CC BY 4.0).}
\endgroup

\begin{abstract}
Large language models used for code editing can be trained and deployed in at least two
distinct output regimes: \emph{direct} generation, where the model emits the entire
modified file in one shot, and \emph{iterative diff-based} generation (``steps''), where
the model emits a sequence of localized search/replace edits that are applied one at a
time until the model signals completion or a step budget is exhausted. The diff-based
regime is attractive because it mirrors how human developers edit code and because, in
principle, it should require the model to generate far fewer tokens per turn. We train
two code models --- a 100M-parameter model trained from scratch (Rainbow-Pony-100M) and a
fine-tuned Qwen2.5-Coder-0.5B \citep{qwen25coder} --- in both regimes on a shared
Flutter/Dart code-editing dataset, and evaluate all four resulting models
(architecture~$\times$~regime) on a held-out set of $\approx$1{,}790 tasks per model. We
find that direct generation substantially outperforms iterative diff-based generation on
every metric we measure --- compilation/static-analysis pass rate, bits-per-byte,
character-level similarity to the reference, and blinded LLM-judge ratings of goal
fulfillment, correctness, and code quality --- and that this gap persists even after
controlling for task difficulty via a matched-ID comparison and even when restricting the
comparison to code that compiles on both sides. We then look for the conditions under
which the diff-based model \emph{does} win, and find a single, architecture-independent
mechanism: diff-based generation is competitive specifically on short, spatially localized
edits (few required edit steps), and its category-level wins concentrate in exactly the
two task categories --- refactoring and error-handling/edge-case fixes --- that
independently have the lowest mean edit-step count in our dataset. We term this
\emph{task locality} and discuss its implications for when an edit-based training regime
is and is not the right choice for a code-editing model.
\end{abstract}

\section{Introduction}
\label{sec:intro}

When a language model is asked to modify an existing source file, there are two natural
ways to have it express the change. The first is to regenerate the entire file from
scratch, conditioned on the original file and the instruction (\emph{direct} generation).
The second is to have the model emit a sequence of localized edits --- typically in a
search/replace or unified-diff format --- that are mechanically applied to the original
file (\emph{iterative} or \emph{diff-based} generation). Production coding agents and
IDE-integrated tools overwhelmingly favor some variant of the second approach
\citep{aider-edit-formats}, for reasons that are intuitive: a diff is shorter than a whole
file, so it is cheaper to generate and less likely to silently corrupt code the model was
never asked to touch; it also mirrors the unit of work a human reviewer actually looks at
(a pull-request diff, not a whole-file rewrite).

Whether this intuition holds up as a \emph{training} objective, rather than just an
\emph{inference-time} interface choice, is less settled. Recent work is mixed. LintSeq
\citep{lintseq} shows that training on synthetic \emph{edit sequences} --- decomposing a
reference program into a chain of small, lint-error-free edits --- improves downstream
code-synthesis quality relative to training on the final program alone, arguing that edit
sequences are a better curriculum, not just a better interface. Conversely, practical
edit-format benchmarks \citep{aider-edit-formats} report that diff-style formats increase
the rate of \emph{malformed} edits (edits the harness cannot apply at all) relative to
whole-file replacement, especially for weaker models, trading a token-efficiency win for a
reliability cost. Adaptive-format work \citep{diff-or-not} goes further and argues that
neither format is uniformly better --- the right format is a property of the specific
edit, and a model (or router) that can choose per-edit outperforms a model committed to
either format across the board.

This paper is an empirical contribution to that question, run end-to-end on our own
models rather than third-party benchmark leaderboards, on a single narrow but realistic
domain: Flutter/Dart code editing. We train two architecturally very different models ---
a small transformer trained from scratch and a mid-size pretrained model fine-tuned for
the task --- each in both a \emph{direct} and a \emph{steps} (iterative diff-based)
variant, holding the training data, tokenization pipeline, and evaluation harness fixed
across all four resulting models. This within-domain, within-dataset design lets us
isolate the effect of the output regime itself from the many confounds (different base
models, different datasets, different edit-format syntax) that make cross-paper
comparisons on this question difficult.

\paragraph{Contributions.} We report four findings:
\begin{enumerate}[leftmargin=*]
    \item Across both architectures, direct generation beats iterative diff-based
    generation by a wide margin on every metric we measure (Section~\ref{sec:aggregate}),
    and the gap is \emph{not} explained away by the diff-based model running out of its
    edit-step budget or by outright edit-application failures --- the majority of
    diff-based failures occur in trajectories that completed normally
    (Section~\ref{sec:attribution}).
    \item The gap survives a matched-ID comparison that controls for the possibility that
    diff-based failures concentrate on intrinsically harder tasks
    (Section~\ref{sec:matched}), and survives restricting the comparison to only the code
    that compiles on both sides, as independently confirmed by a blinded LLM judge scoring
    goal fulfillment, correctness, and code quality (Section~\ref{sec:judge}).
    \item Despite the aggregate gap, there is a real and reproducible subset of tasks
    where the diff-based model wins on judge-rated quality, and this subset is not random:
    it concentrates heavily in short edit trajectories (Section~\ref{sec:wins}).
    \item The category- and trajectory-length-based findings are not two separate
    phenomena: the two task categories where diff-based wins are overrepresented
    (refactoring and error-handling/edge-case fixes) are, independently, the two lowest
    mean-edit-step-count categories in the dataset for both architectures. We unify these
    into a single explanatory variable we call \emph{task locality}
    (Section~\ref{sec:locality}).
\end{enumerate}

\section{Related Work}
\label{sec:related}

\paragraph{Edit formats as an inference-time interface.} The most direct practical
precedent for this work is the Aider project's ongoing benchmarking of edit formats
across many LLMs \citep{aider-edit-formats}, which finds that diff-style formats
(unified diff, search/replace) reduce token cost relative to whole-file replacement but
increase the incidence of edits the harness cannot mechanically apply, with the effect
more pronounced in weaker models. Our \texttt{apply\_failed} and \texttt{malformed}
\texttt{stop\_reason} categories (Section~\ref{sec:harness}) are a direct analogue of this
failure mode, measured end-to-end on models we trained and evaluated ourselves rather than
via API calls to third-party models.

\paragraph{Diff/edit sequences as a training curriculum.} LintSeq \citep{lintseq}
decomposes reference programs into synthetic, lint-clean edit sequences and shows that
training \emph{on the sequence} (not just the final program) improves pass@1 on
downstream code-synthesis benchmarks relative to training on final programs alone. Our
``steps'' models are trained on edit trajectories in a similar spirit, but applied to a
code-\emph{editing} task (modify an existing file) rather than code synthesis from a blank
slate, and our result is directionally opposite for this task/domain: on Flutter/Dart
editing specifically, the edit-trained model underperforms the direct model by a wide
margin. We view these as reconcilable rather than contradictory --- see
Section~\ref{sec:discussion} for discussion of why editing an existing, semantically
constrained file may behave differently from synthesizing a new one.

\paragraph{Adaptive and hybrid formats.} \citet{diff-or-not} argue that the choice between
diff and whole-file output should be adaptive per-edit rather than fixed per-model, and
report that different edit categories favor different formats. Our task-locality finding
(Section~\ref{sec:locality}) is complementary evidence for exactly this claim, obtained
independently on a different domain: we find that the diff-based model is specifically
competitive on short, spatially localized edits and specifically weak on longer,
non-local ones, which is consistent with an adaptive-format model doing better than either
fixed-format model.

\paragraph{Iterative program repair and multi-step editing agents.} Iterative,
multi-turn editing is also the dominant paradigm in LLM coding agents evaluated on
SWE-bench-style benchmarks \citep{swebench-pro, swe-edit, art-of-repair}, where a model
proposes a patch, observes tool/test feedback, and revises. Our steps mode is a simpler,
single-pass version of this loop (no execution feedback between edits; the model commits
to a full edit trajectory conditioned only on its own prior edits), and our results should
not be read as a claim about agentic, feedback-driven repair loops, which are architected
differently and evaluated on a different distribution of tasks (bug localization and
repair in large real-world repositories, rather than small self-contained Flutter/Dart
snippets). We discuss this scoping limitation in Section~\ref{sec:limitations}.

\paragraph{Instruction-tuned code editing.} \citet{instr-tune-code-edit} study
instruction-tuning specifically for code-editing tasks and report that response format
choices materially affect edit quality, again consistent with edit format being a real
design axis rather than a purely cosmetic one.

\section{Methodology}
\label{sec:method}

\subsection{Models and training regimes}
\label{sec:models}

We use two backbones with very different capacity and pretraining history:

\begin{itemize}[leftmargin=*]
    \item \textbf{Rainbow-Pony-100M}: a $\sim$100M-parameter decoder-only transformer
    trained entirely from scratch for this project. The shared pretrained checkpoint
    (before either direct- or steps-mode fine-tuning) is released as
    \texttt{bbidpa/Rainbow-Pony-100m-Flutter-base}: pretrained for 119{,}000 steps
    (batch size 16, block size 1{,}024; 16{,}384 tokens/step) on $\approx$1.95 billion
    tokens --- $\approx$0.79 of one epoch over a 2.60-billion-token corpus (2.47B train
    / 131M validation tokens; 70\% Flutter/Dart source code, 30\% English text) --- using
    a custom 16k-vocabulary BPE tokenizer, with a cosine learning-rate schedule peaking
    at $3\times10^{-4}$ and decaying to $2\times10^{-5}$ by the final logged step, and
    final train/validation loss of 1.28/1.31 (last logged evaluation, step 118{,}800).
    The pretrain-stage checkpoint has
    98{,}146{,}432 parameters; resizing the vocabulary to 16{,}022 post-hoc to
    accommodate structural special tokens such as \texttt{<GOAL>}, \texttt{<CODE>}, and
    the steps-mode action tags brings the two fine-tuned checkpoints to 98{,}163{,}350
    parameters each. Unlike
    Qwen2.5-Coder, this backbone has no exposure to any other programming language or to
    a general-purpose multi-language code pretraining corpus, which isolates the effect
    of output regime from any confound introduced by a broadly-pretrained backbone's own
    biases toward one format or another.
    \item \textbf{Qwen2.5-Coder-0.5B} \citep{qwen25coder}: a pretrained code model
    ($\sim$0.5B parameters, itself derived from Qwen2.5-0.5B), fine-tuned on the same
    underlying Flutter/Dart task pool as Rainbow-Pony (Section~\ref{sec:trainingdata}).
    This tests whether the direct-vs-steps gap is an artifact of an undertrained
    from-scratch model or persists in a model that already has substantial
    code-generation prior.
\end{itemize}

Each backbone is fine-tuned in two regimes on task data derived from the \emph{same}
underlying pool of source examples (Section~\ref{sec:trainingdata}):

\begin{itemize}[leftmargin=*]
    \item \textbf{direct}: given the initial file and an edit instruction, the model
    generates the complete modified file in a single forward pass.
    \item \textbf{steps}: given the initial file and instruction, the model generates a
    sequence of search/replace edit actions. Each action is mechanically applied to the
    current file state (see \texttt{apply\_edit} below) before the next action is
    generated, until the model emits an explicit stop action or a maximum step budget
    (20 steps) is reached.
\end{itemize}

This yields four arms --- \texttt{rainbow-pony-direct}, \texttt{rainbow-pony-steps},
\texttt{qwen-direct}, and \texttt{qwen-steps} --- all evaluated on the same held-out
Flutter/Dart task set under greedy decoding.

\subsubsection{Training data}
\label{sec:trainingdata}

Both fine-tuning datasets derive from the same underlying pool of 14{,}600
hand-designed Flutter/Dart tasks, released as \texttt{bbidpa/flutter-full-examples-v1}
(\texttt{goal}, \texttt{initial\_code}, \texttt{final\_code} triples spanning 36 task
types across three complexity tiers; Apache-2.0 license). This pool is used directly as
the direct-mode fine-tuning data (5M tokens sampled from it) and is also the source for
a step-decomposition procedure --- in the spirit of LintSeq's synthetic edit sequences
\citep{lintseq} --- that expands each full-file example into a forward/backward
sequence of individual search/replace edits, released as
\texttt{bbidpa/flutter-diff-steps-v1} (100K--1M rows, each row one step in a
trajectory linked back to its source example via \texttt{source\_example\_id};
Apache-2.0 license). Steps-mode fine-tuning draws 50M tokens from this decomposed set.

We flag explicitly that this means the two fine-tuning regimes are \emph{not}
token-matched: steps mode receives roughly $10\times$ more fine-tuning tokens than
direct mode (50M vs.\ 5M), a direct consequence of a single full-file example expanding
into many step-level training rows under decomposition. We discuss the implication of
this asymmetry in Section~\ref{sec:limitations}.

Table~\ref{tab:trainingconfig} reports the fine-tuning configuration actually used for
each arm. All four arms share a batch size of 8 and a block size of 1{,}024. The
direct-mode arms were each trained for 5{,}000 steps ($\approx$3.0 target epochs over
their respective $\approx$13.1k-example train splits); the steps-mode arms were each
trained for 43{,}000 steps ($\approx$3.0 target epochs over their respective
step-decomposed train splits). Three of the four arms ---
\texttt{rainbow-pony-direct}, \texttt{rainbow-pony-steps}, and \texttt{qwen-steps} ---
used a cosine learning-rate schedule with peak LR $3\times10^{-5}$ decaying to a floor
of $3\times10^{-6}$, confirmed directly from the full per-step training logs: each of
these three runs' logged LR reaches $3.00\times10^{-5}$ shortly after warmup (step
$\sim$500) and decays to $3.0\times10^{-6}$ at its final logged step.

\texttt{qwen-direct} did \emph{not} follow this schedule, and we disclose this rather
than silently correct it after the fact. Its cosine scheduler was built while the
run's target step count was, at that point in our training script, still set to the
step-mode value (43{,}000) rather than the 5{,}000 steps this arm actually ran, so the
scheduler's decay horizon was roughly $8.5\times$ longer than the run itself: at step
5{,}000 the schedule had only traversed the first $\sim$12\% of its intended cosine
decay. The logged LR for this run confirms this exactly --- it reaches
$3.00\times10^{-5}$ after warmup as intended, but only decays to
$2.9$--$3.0\times10^{-5}$ (not $10^{-6}$) by its final logged step, roughly an order of
magnitude higher than the other three arms at the same point in training. All
\texttt{qwen-direct} results reported in this paper are computed from the checkpoint
this run actually produced; see Section~\ref{sec:limitations} for discussion of the
likely direction of this deviation's effect. Final train/validation loss, taken from
each run's last logged evaluation, is available for all four arms.

\begin{table}[htbp]
\centering
\caption{Fine-tuning configuration and outcome per arm, as logged by the training
harness. Batch size 8 and block size 1{,}024 throughout.}
\label{tab:trainingconfig}
\small
\begin{tabular}{lrrrrrr}
\toprule
Arm & Params & Steps & Epochs & Fine-tune tokens & Final train loss & Final val loss \\
\midrule
rainbow-pony-direct & 98.16M & 5{,}000 & 3.06 & 5.03M / 0.54M (val) & 0.157 & 0.189 \\
rainbow-pony-steps  & 98.16M & 43{,}000 & 2.99 & 146.1M processed$^{*}$ & 0.221 & 0.292 \\
qwen-direct         & 493.81M & 5{,}000 & 3.05 & 4.71M / 0.52M (val) & 0.074 & 0.162 \\
qwen-steps          & 493.81M & 43{,}000 & 2.96 & 138.3M processed$^{*}$ & 0.202 & 0.247 \\
\bottomrule
\end{tabular}
\end{table}
\noindent $^{*}$The steps-mode fine-tuning \emph{dataset} is $\approx$50M tokens
(Section~\ref{sec:trainingdata}); ``tokens processed'' is larger because training ran
for $\approx$3 epochs over it, re-visiting the same tokens multiple times, whereas the
direct-mode dataset column reports the (single-epoch-sized) token count of the
underlying train/validation split directly.

\subsubsection{Action format}
\label{sec:actionformat}

Concretely, each steps-mode instance is a plain-text prompt with tagged sections ---
\texttt{<GOAL>}, \texttt{<CODE>} (the file's current state), \texttt{<HISTORY>} (prior
actions in the trajectory so far, empty on the first step), and an open \texttt{<OUTPUT>}
tag the model completes --- and the model's completion is one \texttt{<ACTION>} record
(an action \texttt{TYPE} and a natural-language \texttt{DESC}) plus one or more
\texttt{<CHANGES>} hunks, terminated by \texttt{</OUTPUT>} (with an additional
\texttt{<DONE></DONE>} marker when the action is the trajectory's final step).
Listing~\ref{lst:steps-example} is a real, unedited training instance from
\texttt{flutter-diff-steps-v1} (second step of a two-step trajectory, rendered by the
same \texttt{render\_step\_prompt} function used at both training and inference time):

\begin{lstlisting}[caption={A real steps-mode training instance: prompt (everything through the open \texttt{<OUTPUT>} tag) plus the model's target completion.}, label={lst:steps-example}]
<GOAL>
Create a simple stateless Flutter widget displaying centered text
</GOAL>

<CODE>
import 'package:flutter/material.dart';

</CODE>

<HISTORY>
<ACTION><TYPE>add_import</TYPE><DESC>Added the import statement 'import 'package:flutter/material.dart';'</DESC></ACTION>
</HISTORY>

<OUTPUT>
<ACTION><TYPE>add_method</TYPE><DESC>Added the main method with runApp call.</DESC></ACTION>
<CHANGES>
<HUNK>
<SEARCH>
import 'package:flutter/material.dart';

</SEARCH>
<REPLACE>
import 'package:flutter/material.dart';

void main() {
  runApp(MaterialApp(home: SimpleTextWidget()));
}

</REPLACE>
</HUNK>
</CHANGES>
</OUTPUT>
\end{lstlisting}

Direct mode uses the same \texttt{<GOAL>}, \texttt{<CODE>}, and \texttt{<OUTPUT>} tags but
no \texttt{<HISTORY>} or \texttt{<ACTION>}\slash\texttt{<CHANGES>} structure: the model's
completion is simply the complete modified (or newly created) file, verbatim, followed by
\texttt{</OUTPUT>}.

\subsubsection{Edit application and the fallback heuristic}
\label{sec:apply}

Each steps-mode edit action specifies a \texttt{search} span and a \texttt{replace} span.
Application is exact-match: if \texttt{search} occurs in the current file exactly once, it
is replaced; if it occurs zero times, the edit is rejected outright (an
\texttt{apply\_failed} step). If it occurs \emph{more than once}, a fallback heuristic is
invoked to disambiguate, since rejecting on ambiguity alone would make every
short-and-generic search span an automatic failure. Our fallback resolves ambiguity by
matching the \emph{first} occurrence of \texttt{search} in the file:

\begin{quote}
\ttfamily
\begin{tabular}{l}
count = code.count(search) \\
if count $\le$ 1: raise (re-raised as apply\_failed) \\
index = code.find(search) \\
return code[:index] + replace + code[index + len(search):]
\end{tabular}
\end{quote}

We flag this as a heuristic, not a principled fix: neither first-occurrence nor
last-occurrence resolution is reliably correct in general, since either can silently edit
the wrong instance of a repeated span. A more robust design would reject overly generic or
short \texttt{search} blocks outright rather than guessing; we did not implement this and
note it as a source of some of the \texttt{done}-but-still-\texttt{dart\_pass=False}
trajectories discussed in Section~\ref{sec:fallback}.

\subsection{Evaluation harness and metrics}
\label{sec:harness}

Each model is evaluated on the same $\approx$1{,}790-example held-out set
(\texttt{rainbow-pony}: $n=1{,}789$; \texttt{qwen}: $n=1{,}792$; the small difference is
attributable to differing tokenizer behavior under a fixed 1024-token block size during
tokenization, which drops a handful of examples differently per tokenizer). For each
example we record:

\begin{itemize}[leftmargin=*]
    \item \textbf{dart\_pass}: whether the model's final output passes Dart static
    analysis (\texttt{dart analyze}) --- our primary correctness signal.
    \item \textbf{bits\_per\_byte}: model perplexity on the reference completion,
    computed identically regardless of mode (a single-shot teacher-forced score against
    the reference \texttt{final\_code}, decoupled from the steps trajectory itself ---
    see the caveat in Section~\ref{sec:winrates}).
    \item \textbf{similarity\_ratio}: character-level similarity between the model's
    output and the reference \texttt{final\_code}.
    \item \textbf{stop\_reason} (steps mode only): why the trajectory ended ---
    \texttt{done} (model emitted an explicit stop action), \texttt{max\_steps} (20-step
    budget exhausted), \texttt{apply\_failed} (an edit could not be applied and no
    fallback rescued it), or \texttt{malformed} (the model emitted an unparseable
    action).
    \item \textbf{num\_steps}, \textbf{num\_fallback\_steps} (steps mode only): the
    trajectory length and how many of those steps required the ambiguity fallback of
    Section~\ref{sec:apply}.
\end{itemize}

\subsection{Matched-ID (``clean'' / ``best-case'') comparison}
\label{sec:matched-method}

A naive direct-vs-steps comparison on the full held-out set risks conflating two distinct
effects: (a) steps mode is worse \emph{at the same task}, and (b) steps mode's failures
happen to concentrate on tasks that are independently harder. To separate these, we define
a \emph{clean} steps-mode subset per architecture:
\begin{small}
\[
\begin{aligned}
\texttt{clean\_steps} = \{\, r \in \texttt{steps\_results} : \; & \texttt{stop\_reason}(r) = \texttt{done} \\
& \;\wedge\; \texttt{num\_fallback\_steps}(r) = 0 \;\wedge\; \texttt{num\_steps}(r) < 20 \,\}
\end{aligned}
\]
\end{small}
i.e.\ trajectories that completed normally, required no ambiguity fallback, and did not
hit the step budget. We then compare this subset against the \emph{direct}-mode results on
the \emph{same sample IDs} (\texttt{matched\_direct}), rather than against direct's full
results. We refer to this restricted, same-ID comparison as the \emph{matched-ID}
comparison throughout, and to it further restricted to \texttt{dart\_pass=True} steps rows
as the \emph{best-case} comparison, since it isolates the specific population of steps
trajectories a practitioner could hope to reach with more training (no fallback, no
budget exhaustion, and a correct result).

\subsection{Blinded LLM-as-judge protocol}
\label{sec:judge-method}

To validate that \texttt{dart\_pass} (a binary static-analysis signal) is not masking
quality differences among code that compiles on \emph{both} sides, we additionally score a
subset of outputs with an LLM judge. The judge is shown only the task instruction, the
initial file, and a single candidate output file; it is never told the model name,
training mode, or the \texttt{dart\_pass} outcome for that candidate (a fully blinded,
single-candidate protocol --- the judge scores one output in isolation per call, not a
head-to-head pair). It returns three integer ratings on a 1--5 scale via a
structured-output schema: \texttt{goal\_fulfillment}, \texttt{correctness}, and
\texttt{code\_quality}, plus a free-text justification. The judge model used throughout
is \texttt{gpt-4.1}. Unlike a subsampled audit, the judge was run over essentially the
full held-out set for all four arms: 1,792/1,792 rows for \texttt{qwen-direct} and
\texttt{qwen-steps}, 1,789/1,789 for \texttt{rainbow-pony-direct}, and 1,788/1,789 for
\texttt{rainbow-pony-steps} (one row skipped with \texttt{judge\_error=empty\_output\_code},
i.e.\ the model produced no output to score) --- 7,161 judged rows in total across the
four evaluation datasets.

\section{Results}
\label{sec:results}

\subsection{Aggregate performance}
\label{sec:aggregate}

Table~\ref{tab:aggregate} summarizes all three core metrics across the four arms.
Wilson 95\% confidence intervals are reported for \texttt{dart\_pass}.

\begin{table}[htbp]
\centering
\caption{Aggregate evaluation metrics by architecture and mode. $n$ is the held-out set
size per architecture (shared across its direct/steps arms).}
\label{tab:aggregate}
\begin{tabular}{llrccc}
\toprule
Architecture & Mode & $n$ & dart\_pass (95\% CI) & bits/byte & similarity\\
\midrule
Rainbow-Pony-100M & direct & 1{,}789 & 0.802 [0.783, 0.820] & 0.107 & 0.511 \\
Rainbow-Pony-100M & steps  & 1{,}789 & 0.347 [0.325, 0.369] & 0.180 & 0.439 \\
Qwen2.5-Coder-0.5B & direct & 1{,}792 & 0.900 [0.885, 0.913] & 0.088 & 0.568 \\
Qwen2.5-Coder-0.5B & steps  & 1{,}792 & 0.501 [0.478, 0.524] & 0.142 & 0.493 \\
\bottomrule
\end{tabular}
\end{table}

Direct generation beats steps-mode generation by a wide, non-overlapping margin on
\texttt{dart\_pass} for both architectures (45.5 percentage points for Rainbow-Pony;
39.9 points for Qwen), and consistently on \texttt{bits\_per\_byte} and
\texttt{similarity\_ratio} as well. The gap is present regardless of whether the backbone
was trained from scratch or fine-tuned from a strong pretrained checkpoint, which argues
against ``the from-scratch model just hasn't learned the edit format yet'' as a complete
explanation. 

\begin{figure}[htbp]
\centering
\includegraphics[width=0.72\textwidth,height=0.85\textheight,keepaspectratio]{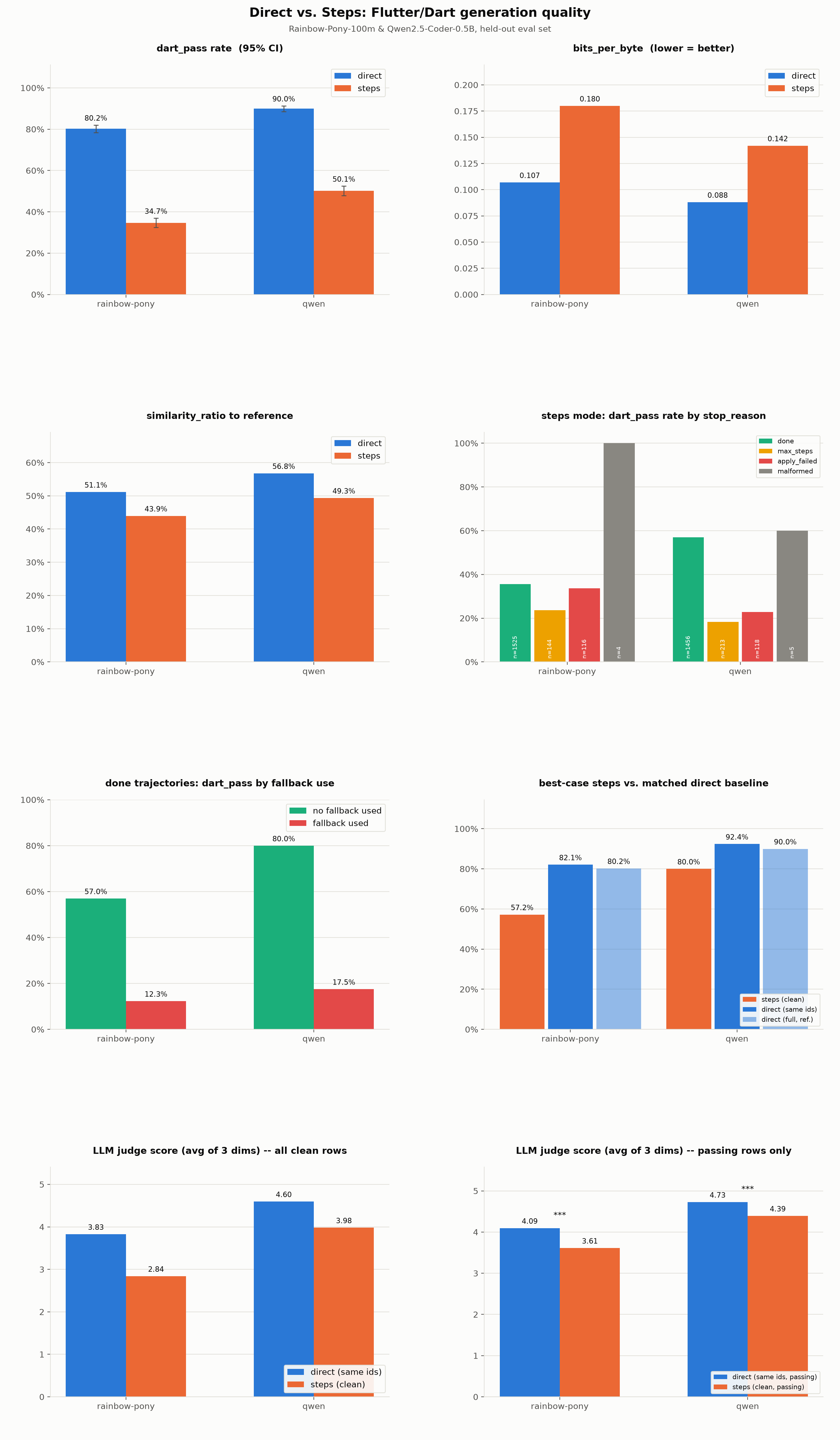}
\caption{Summary panel: \texttt{dart\_pass} rate, \texttt{bits\_per\_byte},
\texttt{similarity\_ratio}, stop-reason and fallback breakdowns, the matched-ID
best-case comparison, and blinded judge scores, across both architectures.}
\label{fig:summary}
\end{figure}

Figure~\ref{fig:summary} visualizes these results together with the stop-reason,
fallback, and judge-score breakdowns discussed in the remainder of this section.

\subsection{Failure attribution: where do steps-mode failures come from?}
\label{sec:attribution}

A natural hypothesis is that steps-mode's lower \texttt{dart\_pass} rate is mostly a
process-failure artifact: the model runs out of its step budget, or an edit cannot be
applied. Table~\ref{tab:stopreason} shows this is not the dominant story.

\begin{table}[htbp]
\centering
\caption{Steps-mode trajectory outcomes (\texttt{stop\_reason}) by architecture.}
\label{tab:stopreason}
\begin{tabular}{lrrrr}
\toprule
Architecture & done & apply\_failed & max\_steps & malformed \\
\midrule
Rainbow-Pony-100M & 1{,}525 (85.2\%) & 116 (6.5\%) & 144 (8.1\%) & 4 (0.2\%) \\
Qwen2.5-Coder-0.5B & 1{,}456 (81.3\%) & 118 (6.6\%) & 213 (11.9\%) & 5 (0.3\%) \\
\bottomrule
\end{tabular}
\end{table}

The large majority of steps-mode trajectories (81--85\%) complete normally
(\texttt{stop\_reason=done}) rather than failing outright. Table~\ref{tab:dartbyreason}
breaks \texttt{dart\_pass} down by \texttt{stop\_reason}.

\begin{table}[htbp]
\centering
\caption{\texttt{dart\_pass} rate conditional on \texttt{stop\_reason}, steps mode.
\texttt{malformed} has $n\le5$ per architecture and is noise.}
\label{tab:dartbyreason}
\begin{tabular}{lrrrr}
\toprule
Architecture & done & apply\_failed & max\_steps & malformed ($n$) \\
\midrule
Rainbow-Pony-100M & 0.356 & 0.336 & 0.236 & 1.00 (4) \\
Qwen2.5-Coder-0.5B & 0.569 & 0.229 & 0.183 & 0.60 (5) \\
\bottomrule
\end{tabular}
\end{table}

Because \texttt{done} trajectories are both the large majority of the data \emph{and} have
a \texttt{dart\_pass} rate well below the direct-mode baseline, they account for the bulk
of total steps-mode failures: approximately 84\% of all Rainbow-Pony steps-mode failures
and approximately 70\% of all Qwen steps-mode failures occur in trajectories that
completed normally, not in trajectories that hit the step budget or suffered an
unrecoverable apply failure. In other words, most of the quality gap is \emph{silent
content corruption within successfully-completed trajectories}, not budget exhaustion or
mechanical apply errors.

\subsection{The role of fallback edits}
\label{sec:fallback}

Section~\ref{sec:apply} noted that ambiguous \texttt{search} spans are resolved via a
first-occurrence heuristic rather than rejected. Table~\ref{tab:fallback} shows this
heuristic is a major driver of failure \emph{even among} \texttt{done} trajectories.

\begin{table}[htbp]
\centering
\caption{\texttt{dart\_pass} rate within \texttt{done}-only trajectories, split by
whether the trajectory required at least one ambiguity-fallback edit.}
\label{tab:fallback}
\begin{tabular}{lrr}
\toprule
Architecture & no fallback ($n$) & fallback used ($n$) \\
\midrule
Rainbow-Pony-100M & 0.570 (795) & 0.123 (730) \\
Qwen2.5-Coder-0.5B & 0.800 (919) & 0.175 (537) \\
\bottomrule
\end{tabular}
\end{table}

Trajectories that required at least one fallback resolution pass at less than a quarter of
the rate of trajectories that never needed one, for both architectures. Despite being a
minority of \texttt{done} rows, fallback-affected trajectories account for roughly 65\%
(Rainbow-Pony) and 71\% (Qwen) of all failures within the \texttt{done} bucket. This
localizes a large share of the quality gap to a specific, identifiable mechanism: ambiguous
edit targets that the model itself created (by emitting a \texttt{search} span that
matches multiple locations) and that our disambiguation heuristic cannot reliably resolve.

\subsection{Correcting for task-selection bias: matched-ID comparison}
\label{sec:matched}

Applying the matched-ID methodology of Section~\ref{sec:matched-method} to Rainbow-Pony
yields 792 ``clean'' steps IDs (done, no fallback, $<$20 steps). Comparing this subset
against direct-mode results restricted to the \emph{same} IDs (Table~\ref{tab:matched})
shows two things. First, the clean subset is measurably \emph{easier for direct mode too}
(matched-direct's dart\_pass rate exceeds direct's full-population rate from
Table~\ref{tab:aggregate}), confirming that ``clean'' steps trajectories are not a random
sample of tasks --- they skew toward intrinsically easier edits. Second, and more
importantly, once this selection effect is accounted for, the residual gap between steps
and direct is \emph{larger}, not smaller, than the naive full-population gap.

\begin{table}[htbp]
\centering
\caption{Matched-ID comparison: clean steps-mode trajectories vs.\ direct mode on the
identical sample IDs. Recomputed directly from the released evaluation datasets
(Section~\ref{sec:availability}).}
\label{tab:matched}
\begin{tabular}{lrrrr}
\toprule
Architecture & clean IDs ($n$) & clean-steps dart\_pass & matched-direct dart\_pass & gap (pp) \\
\midrule
Rainbow-Pony-100M & 792 & 57.2\% & 82.1\% & 24.9 \\
Qwen2.5-Coder-0.5B & 919 & 80.0\% & 92.4\% & 12.4 \\
\bottomrule
\end{tabular}
\end{table}

\subsection{Independent confirmation via blinded LLM judge}
\label{sec:judge}

The matched-ID comparison still relies on \texttt{dart\_pass}, a binary static-analysis
signal that says nothing about the quality of code that compiles on both sides. We
therefore restrict the matched-ID subset further to rows where \texttt{dart\_pass=True}
on \emph{both} sides, and score those rows with the blinded judge of
Section~\ref{sec:judge-method}. Table~\ref{tab:judge} reports the results.

\begin{table}[htbp]
\centering
\caption{Blinded LLM-judge scores (1--5 scale) among matched-ID rows where both steps and
direct pass \texttt{dart\_pass} ($n=390$ paired rows for Rainbow-Pony-100M, $n=688$ for
Qwen2.5-Coder-0.5B). All six pairwise differences are significant under Welch's
$t$-test at $p<0.001$. Recomputed directly from the released evaluation datasets
(Section~\ref{sec:availability}).}
\label{tab:judge}
\begin{tabular}{llrrrr}
\toprule
Architecture & Dimension & steps & direct & $t$ (Welch) & $p$ \\
\midrule
\multirow{3}{*}{Rainbow-Pony-100M}
 & goal fulfillment & 3.44 & 3.89 & $-4.27$ & $2.2\times10^{-5}$ \\
 & correctness      & 3.71 & 4.14 & $-4.60$ & $5.0\times10^{-6}$ \\
 & code quality      & 3.96 & 4.46 & $-6.87$ & $<10^{-6}$ \\
\midrule
\multirow{3}{*}{Qwen2.5-Coder-0.5B}
 & goal fulfillment & 4.36 & 4.72 & $-6.95$ & $<10^{-6}$ \\
 & correctness      & 4.36 & 4.75 & $-8.17$ & $<10^{-6}$ \\
 & code quality      & 4.52 & 4.82 & $-8.25$ & $<10^{-6}$ \\
\bottomrule
\end{tabular}
\end{table}

Even restricted to code that compiles on both sides, direct-mode output is rated
measurably higher quality by a judge that never saw which arm produced which output. This
is the paper's central negative result for steps mode: roughly half of the raw
\texttt{dart\_pass} gap is explained by steps mode's lower compile rate, but the other
half is a genuine, judge-confirmed residual quality difference among code that compiles
successfully either way.

\begin{figure}[htbp]
\centering
\includegraphics[width=0.85\textwidth]{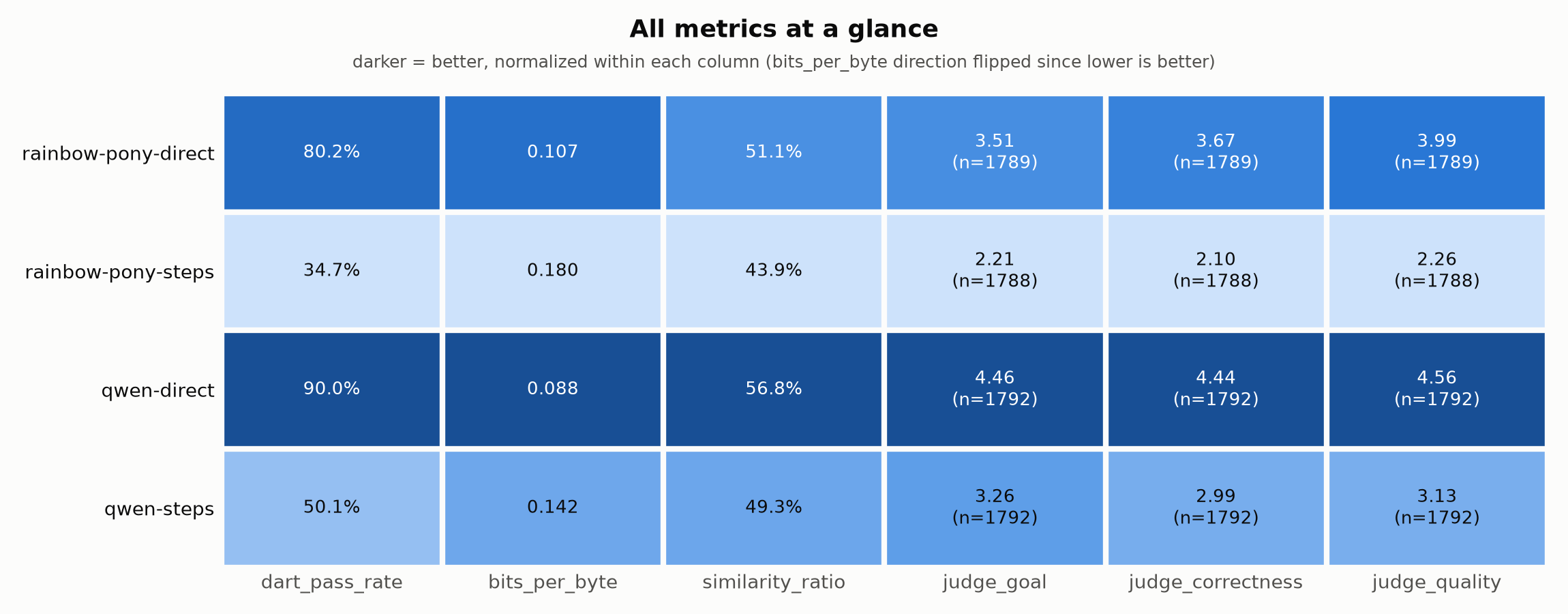}
\caption{Heatmap summary of all four arms across six key metrics (darker = better,
per-column normalized).}
\label{fig:heatmap}
\end{figure}

\begin{figure}[htbp]
\centering
\includegraphics[width=0.85\textwidth]{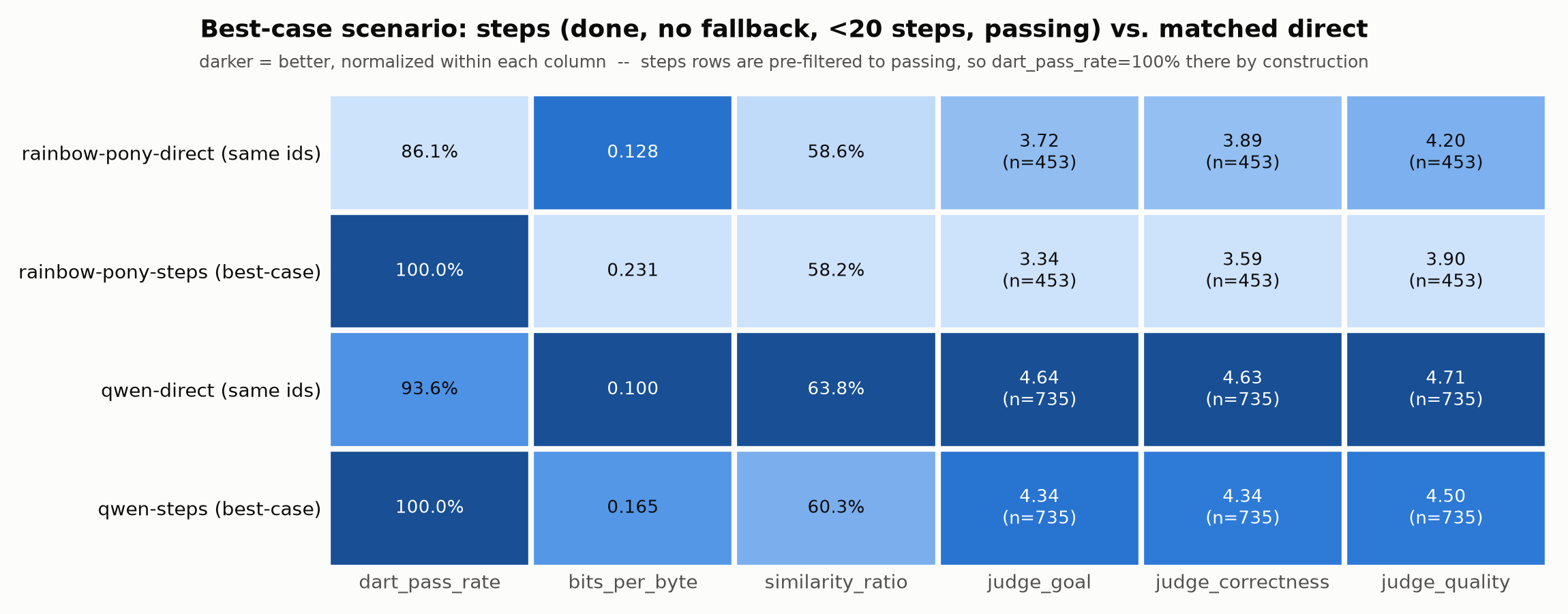}
\caption{Same heatmap restricted to the best-case steps population (done, no fallback,
$<$20 steps, \texttt{dart\_pass=True}) vs.\ matched direct-mode rows.}
\label{fig:heatmap-best}
\end{figure}

Figures~\ref{fig:heatmap} and~\ref{fig:heatmap-best} present the same six-metric
comparison as a heatmap, for the full population and the best-case subset respectively.

\subsection{Where does steps mode win? A paired analysis}
\label{sec:wins}

Sections~\ref{sec:aggregate}--\ref{sec:judge} establish that direct mode wins in
aggregate, robustly. This section asks the complementary question: is there \emph{any}
identifiable subpopulation where steps mode is competitive or better, and if so, what
characterizes it? We answer this via a per-row paired analysis: for every sample ID
present in both a model's direct and steps outputs, we flag whether steps ``wins'' on each
metric (a strict improvement over direct on that same task).

\subsubsection{Per-metric win rates}
\label{sec:winrates}

\begin{table}[htbp]
\centering
\caption{Steps-mode per-row win rate by metric, across both architectures (ranges span
the two architectures).}
\label{tab:winrates}
\begin{tabular}{lc}
\toprule
Metric & steps win rate \\
\midrule
dart\_pass (steps passes, direct fails) & 3.5\% -- 4.8\% \\
bits\_per\_byte (steps lower) & 0.1\% -- 0.3\% \\
similarity\_ratio (steps higher) & 33.7\% -- 34.8\% \\
judge dimensions (steps higher) & 4.8\% -- 9.2\% \\
\bottomrule
\end{tabular}
\end{table}

Table~\ref{tab:winrates} reports the resulting win rate for each metric.

Two of these numbers require a caveat before they can be interpreted as signal.
\texttt{bits\_per\_byte} is computed identically regardless of mode --- a single-shot
teacher-forced score against the reference completion --- so it is structurally decoupled
from whatever the steps trajectory actually did; its near-zero win rate is expected and
uninformative about editing quality. \texttt{similarity\_ratio}'s unusually high win rate
is, we believe, a \emph{metric artifact} rather than a real quality signal: steps mode's
edit-based generation process naturally preserves more character-level overlap with the
reference \texttt{final\_code}, because both were constructed via incremental edits from
the same \texttt{initial\_code}, independent of whether the edits were semantically
correct. We therefore treat the \texttt{dart\_pass} and judge win rates as the meaningful
signal, and focus the remainder of this section on those.

We also tested whether \texttt{initial\_code\_len} (the length of the file being edited)
predicts steps-mode wins, on the hypothesis that longer starting files might favor an
edit-based approach. This hypothesis is rejected: the effect has \emph{opposite sign}
between architectures (Rainbow-Pony wins skew toward 6\% longer initial files; Qwen wins
skew toward 11\% shorter ones), and both effects are small. File length alone does not
explain when steps mode wins.

\subsubsection{Category analysis}
\label{sec:category}

Table~\ref{tab:category} shows the task-category composition of \texttt{dart\_pass}
steps-mode wins relative to each category's baseline share of the dataset (11.2\% under a
9-category uniform baseline).

\begin{table}[htbp]
\centering
\caption{Task categories overrepresented among \texttt{dart\_pass} steps-mode wins,
relative to their $\sim$11.2\% baseline share of the dataset, in both architectures.
Win counts are small ($n=63$--86 total wins per architecture); other categories disagreed
in direction between architectures and are treated as noise.}
\label{tab:category}
\begin{tabular}{lrr}
\toprule
Category & Rainbow-Pony share of wins & Qwen share of wins \\
\midrule
refactoring\_edits & 18.6\% & 14.3\% \\
error\_handling\_and\_edge\_cases & 15.1\% & 15.9\% \\
\midrule
(baseline share, all categories) & \multicolumn{2}{c}{$\sim$11.2\%} \\
\bottomrule
\end{tabular}
\end{table}

These are the only two categories consistently overrepresented among steps-mode wins in
\emph{both} architectures.

\subsubsection{Trajectory length analysis}
\label{sec:steplen}

We next asked whether steps-mode wins concentrate at particular trajectory lengths
(\texttt{num\_steps}). Table~\ref{tab:steplen} compares the mean/median \texttt{num\_steps}
of rows where steps wins on a majority ($\ge2$ of 3) of judge dimensions against all other
rows.

\begin{table}[htbp]
\centering
\caption{Trajectory length (\texttt{num\_steps}) for composite judge-majority wins vs.\
all other rows. All differences significant under Welch's $t$-test, $p\approx0$.}
\label{tab:steplen}
\begin{tabular}{lrrrrr}
\toprule
Architecture & wins mean & wins median & rest mean & rest median & $t$ \\
\midrule
Rainbow-Pony-100M & 5.54 & 5 & 8.94 & 8 & $-8.60$ \\
Qwen2.5-Coder-0.5B & 6.65 & 6 & 9.02 & 8 & $-4.54$ \\
\bottomrule
\end{tabular}
\end{table}

The same pattern holds when the three judge dimensions are tested individually
(Table~\ref{tab:steplen-dims}): in all six tests (2 architectures $\times$ 3 dimensions),
steps-mode wins are associated with significantly shorter trajectories.

\begin{table}[htbp]
\centering
\caption{Per-dimension Welch's $t$-tests: trajectory length of steps-mode judge wins vs.\
non-wins. All $p<0.001$.}
\label{tab:steplen-dims}
\begin{tabular}{llrr}
\toprule
Architecture & Dimension & $t$ & $p$ \\
\midrule
\multirow{3}{*}{Rainbow-Pony-100M}
 & goal fulfillment & $-8.48$ & $<0.001$ \\
 & correctness      & $-9.99$ & $<0.001$ \\
 & code quality      & $-5.09$ & $<0.001$ \\
\midrule
\multirow{3}{*}{Qwen2.5-Coder-0.5B}
 & goal fulfillment & $-3.85$ & $0.0002$ \\
 & correctness      & $-5.39$ & $<0.001$ \\
 & code quality      & $-3.72$ & $0.0003$ \\
\bottomrule
\end{tabular}
\end{table}

\begin{figure}[htbp]
\centering
\includegraphics[width=0.85\textwidth]{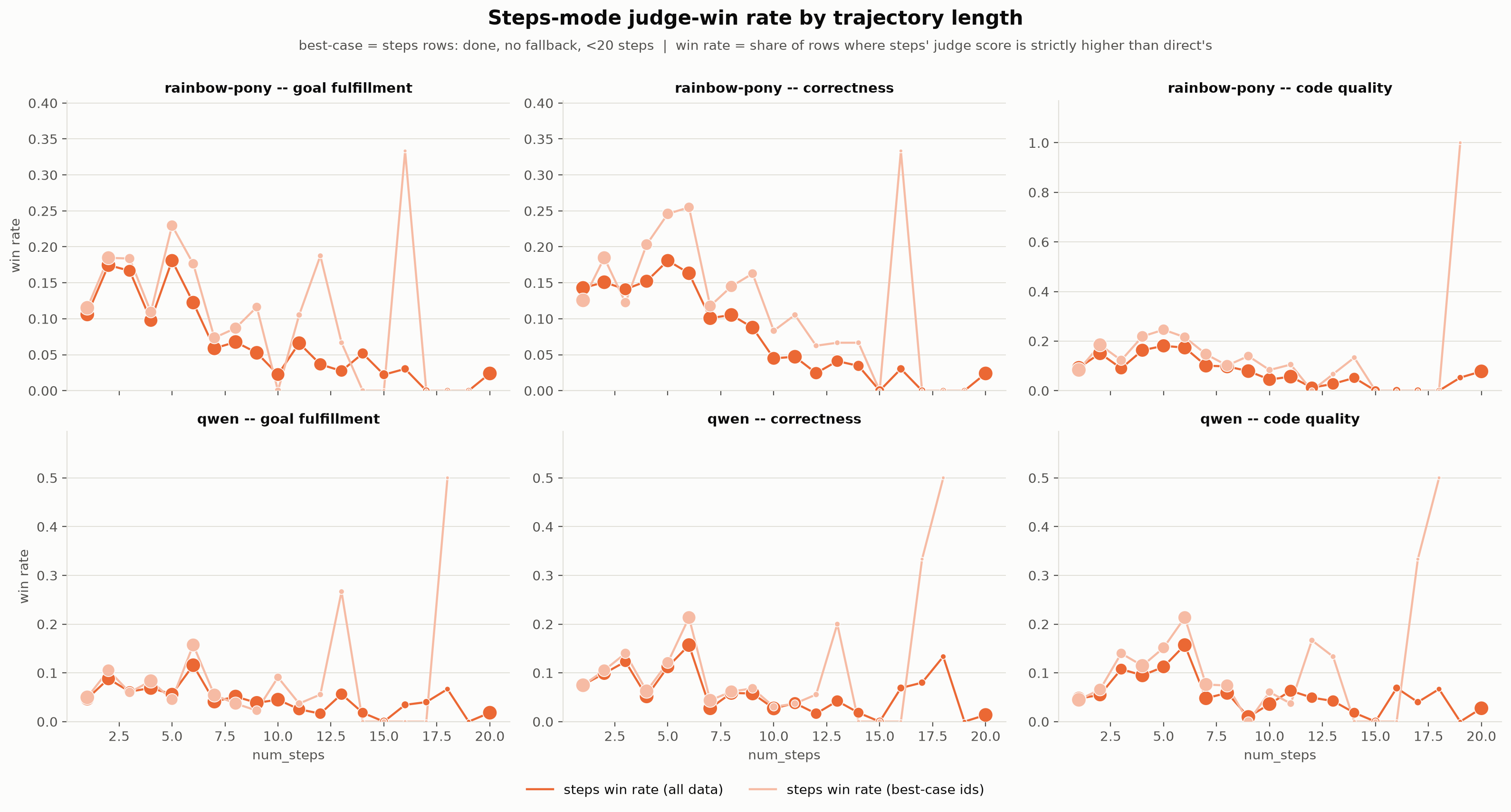}
\caption{Judge win rate by trajectory length (\texttt{num\_steps}), all steps rows vs.\
best-case-only steps rows, per architecture and judge dimension.}
\label{fig:winrate-compare}
\end{figure}

\begin{figure}[htbp]
\centering
\includegraphics[width=0.85\textwidth]{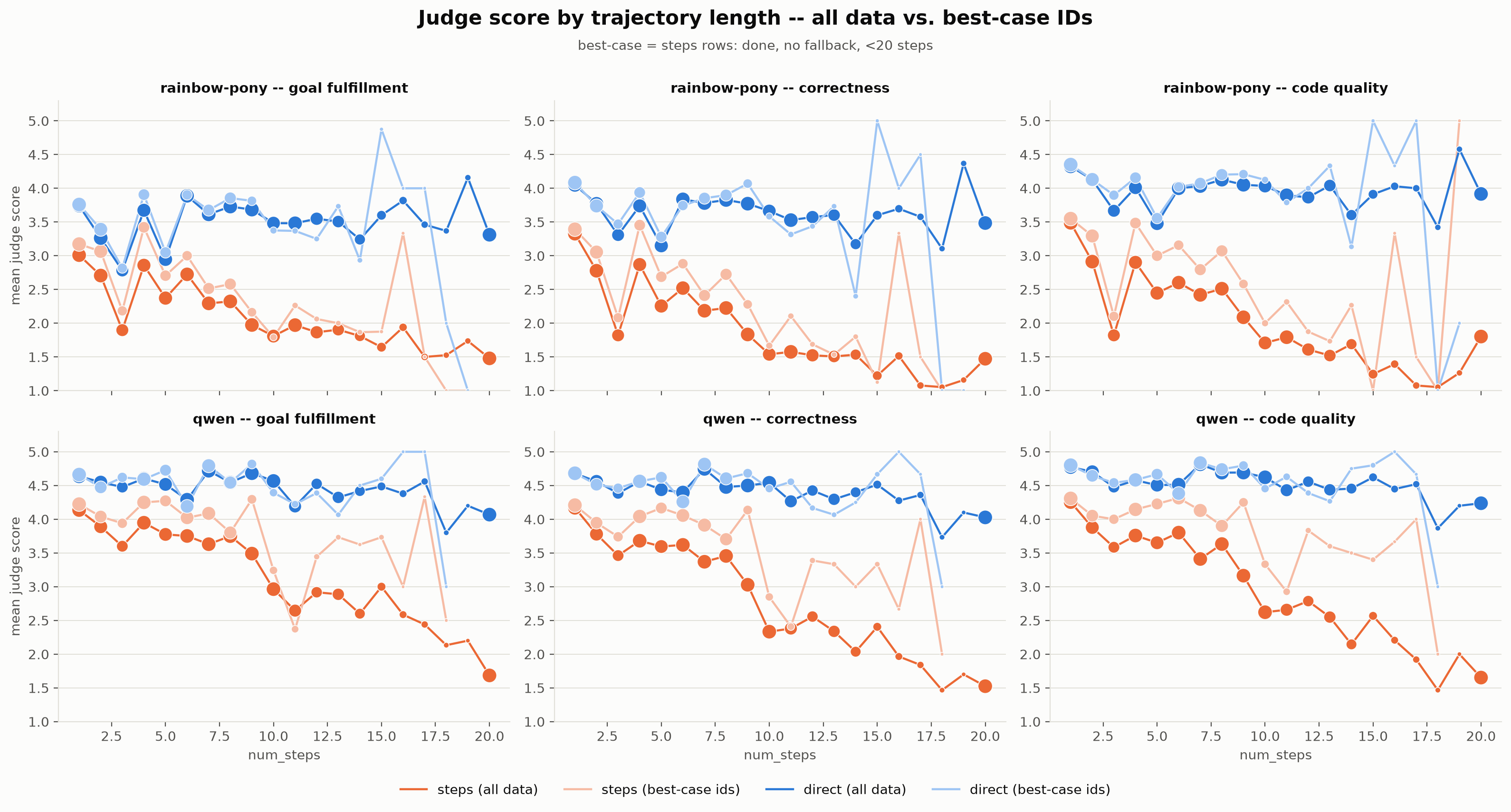}
\caption{Mean judge score by trajectory length: steps (all data), steps (best-case IDs),
direct (all data), direct (best-case IDs).}
\label{fig:fourline}
\end{figure}

\begin{figure}[htbp]
\centering
\includegraphics[width=0.85\textwidth]{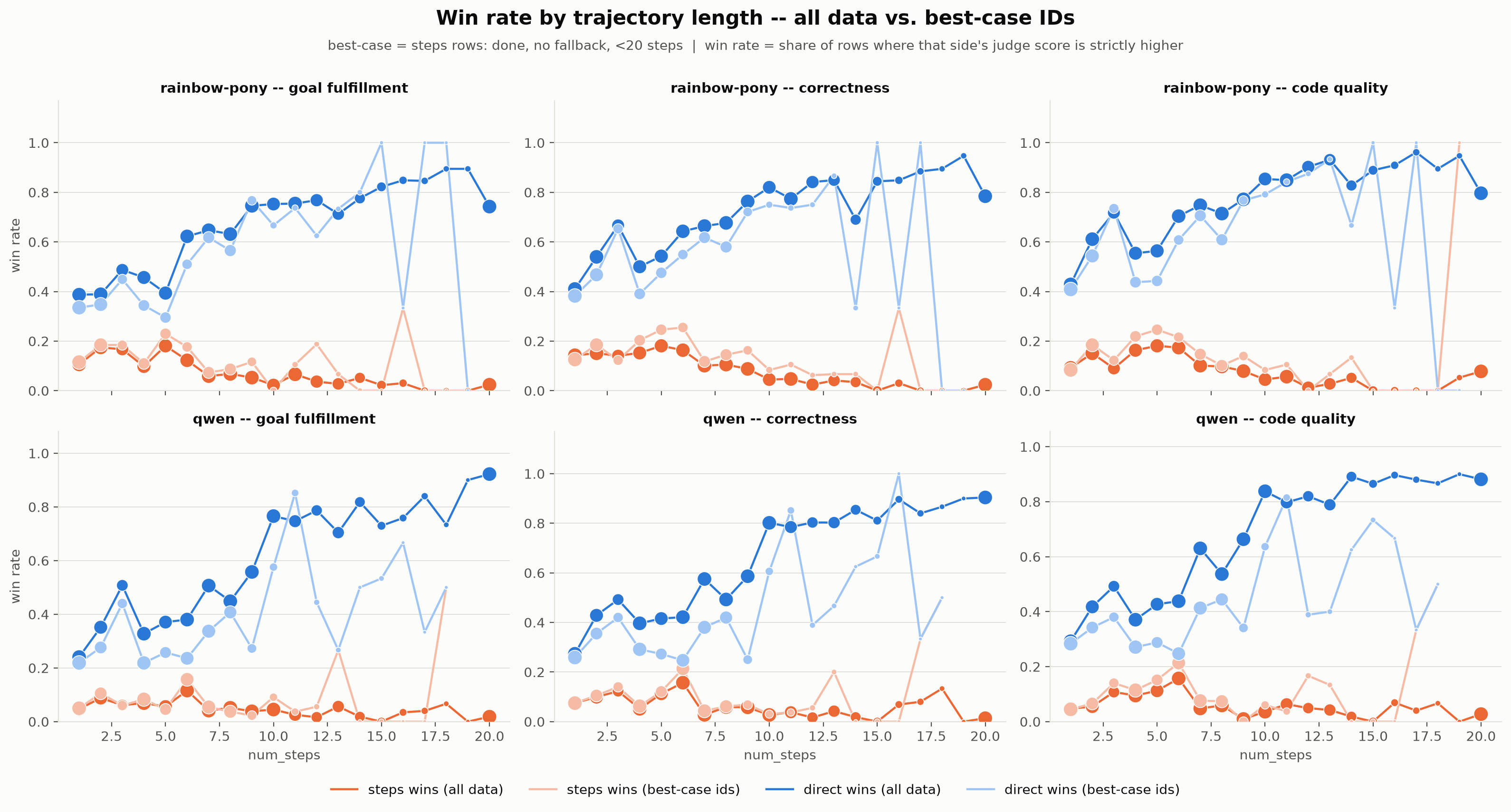}
\caption{Win rate by trajectory length for both directions (steps-wins, direct-wins),
all data vs.\ best-case IDs.}
\label{fig:fourline-winrate}
\end{figure}

\begin{figure}[htbp]
\centering
\includegraphics[width=0.85\textwidth]{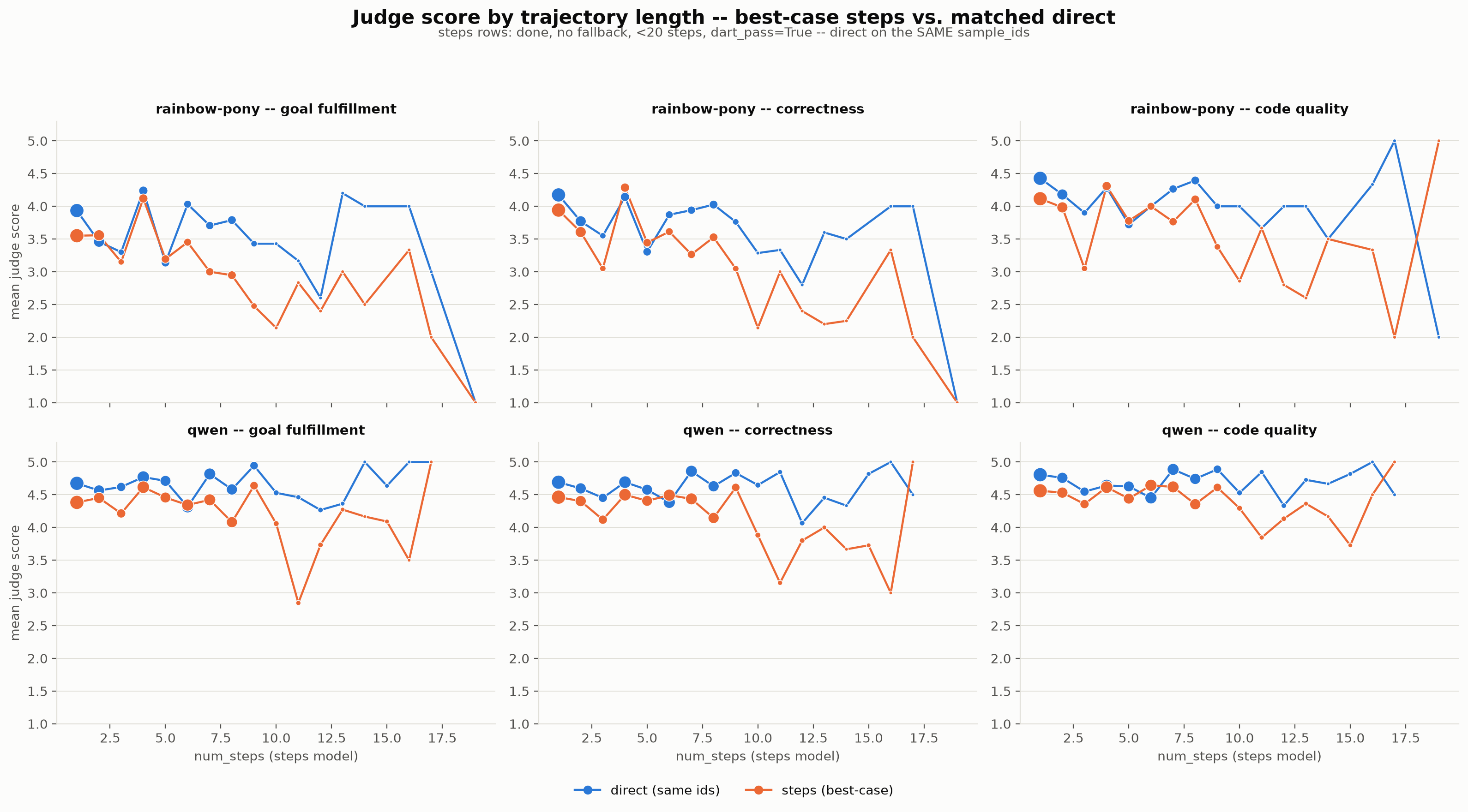}
\caption{Judge score by trajectory length, best-case steps (done, no fallback, $<$20
steps, \texttt{dart\_pass=True}) vs.\ matched direct-mode rows on the same sample IDs.
Shown for reference alongside Figure~\ref{fig:fourline}.}
\label{fig:bestcase-score}
\end{figure}

Win rate declines monotonically (with minor noise) from roughly 10--18\% at
\texttt{num\_steps} 1--6 down to approximately 0\% by \texttt{num\_steps} 14--19, and this
decline is consistent across all six category/architecture breakdowns we examined
(Figures~\ref{fig:winrate-compare} and~\ref{fig:fourline-winrate}). The underlying raw
judge score shows the same pattern from the other direction: steps mode's own score
declines steadily as trajectory length grows, most visibly in Figure~\ref{fig:fourline}
(and Figure~\ref{fig:bestcase-score} for the best-case-only subset), while direct mode's
score on the same rows stays roughly flat across the same range --- the win-rate decline
above is a direct consequence of this asymmetry between the two modes, not an independent
effect.

For reference, Figure~\ref{fig:bestcase-score} shows the same best-case-steps-vs.\
matched-direct judge scores as Figure~\ref{fig:fourline}, but restricted to the
best-case subset on both axes (rather than overlaid against the all-data lines).

\subsection{Unifying the findings: task locality}
\label{sec:locality}

Sections~\ref{sec:category} and~\ref{sec:steplen} report what look like two separate
findings --- a category effect and a trajectory-length effect --- but they are the same
mechanism observed twice. Table~\ref{tab:unify} reports mean \texttt{num\_steps} by
category, sorted ascending, for both architectures.

\begin{table}[htbp]
\centering
\caption{Mean \texttt{num\_steps} for the two win-overrepresented categories vs.\ the
full 9-category distribution. Both are the lowest or near-lowest mean-step-count
categories in \emph{both} architectures.}
\label{tab:unify}
\begin{tabular}{lrr}
\toprule
Category & Rainbow-Pony mean steps & Qwen mean steps \\
\midrule
refactoring\_edits (median 2--3) & 4.14 & 4.49 \\
error\_handling\_and\_edge\_cases & 6.84 & 7.24 \\
\bottomrule
\end{tabular}
\end{table}

\texttt{refactoring\_edits} and \texttt{error\_handling\_and\_edge\_cases} are, in both
architectures, literally the two lowest mean-\texttt{num\_steps} categories out of nine.
This is not a coincidence: it is the same underlying variable driving both the category
finding and the trajectory-length finding. We term this variable \emph{task locality} ---
how spatially narrow and self-contained the required change is --- and conclude that task
locality, not category label or architecture per se, is what predicts when an iterative
diff-based approach is competitive with direct whole-file generation.

\section{Discussion}
\label{sec:discussion}

The aggregate picture is unambiguous: on this Flutter/Dart editing task, at both model
scales we tested, whole-file direct generation is the better default. But the task-locality
finding gives that result a mechanism, not just a verdict. Every additional edit step in a
steps-mode trajectory is an additional opportunity for the ambiguity-fallback heuristic
(Section~\ref{sec:apply}) to silently pick the wrong occurrence of a repeated span, for
error to compound across a longer chain of dependent edits, and for the model to lose
track of file state it is not directly re-reading in full at each step. Direct generation
pays a fixed cost (regenerate everything) regardless of how localized the true change is;
steps mode pays a cost that scales with trajectory length, and our results show that cost
grows faster than its token-efficiency benefit for anything beyond a handful of steps.

This also reframes the relationship to LintSeq's finding \citep{lintseq} that
edit-sequence training improves code \emph{synthesis}. Synthesizing a new program via a
chain of small, individually-verifiable (lint-clean) edits is a fundamentally different
task from \emph{editing an existing, already-correct file} under our steps regime: in
LintSeq's setting, every edit step is checked against a linter before being kept in the
training sequence, and the ``correct'' end state is being constructed incrementally rather
than located inside an already-semantically-constrained file where a wrong disambiguation
can silently break something that was previously working. Our results are consistent with
\citet{diff-or-not}'s adaptive-format argument: the two paradigms are not in conflict once
task locality is taken into account, they are complementary tools for different points on
the same axis.

\section{Limitations}
\label{sec:limitations}

\begin{itemize}[leftmargin=*]
    \item \textbf{Unequal fine-tuning token budgets.} As disclosed in
    Section~\ref{sec:trainingdata}, steps-mode fine-tuning used roughly $10\times$ more
    tokens than direct-mode fine-tuning (50M vs.\ 5M) for both architectures, since
    step-decomposing the same source examples multiplies row count. We did not
    token-match the two regimes. This asymmetry cuts \emph{against} our central finding
    rather than explaining it away: steps mode still underperforms direct mode in
    aggregate despite the larger fine-tuning budget, which if anything strengthens the
    case that the gap reflects something about the output regime itself rather than
    undertraining. We nonetheless flag it as a confound a token-matched follow-up should
    control for directly, particularly for the task-locality result
    (Section~\ref{sec:locality}), where it remains possible that longer trajectories are
    specifically undertrained relative to short ones within the steps-mode fine-tuning
    set.
    \item \textbf{Single-domain evaluation.} All results are on Flutter/Dart code editing
    specifically. We make no claim that the direct-over-steps ranking generalizes to other
    languages or to larger, multi-file repositories.
    \item \textbf{Greedy decoding, single sample.} All four arms are evaluated under
    greedy decoding with one generation per task; we do not measure pass@$k$ or the effect
    of sampling temperature on either regime.
    \item \textbf{No execution feedback in steps mode.} Our steps mode commits to a full
    edit trajectory without observing compiler or test feedback between edits, unlike
    agentic repair loops \citep{swebench-pro,swe-edit,art-of-repair}. Results here should
    not be extrapolated to feedback-driven, tool-using agents.
    \item \textbf{Ambiguity-fallback heuristic.} The first-occurrence fallback
    (Section~\ref{sec:apply}) is a known source of silent errors; a steps-mode model
    trained/evaluated against a stricter apply harness (e.g., one that rejects ambiguous
    \texttt{search} spans outright rather than guessing) might show a smaller gap, since
    some of what we attribute to ``steps mode is worse'' may partly be ``our apply harness
    guesses wrong.''
    \item \textbf{Judge validity.} The blinded LLM judge (\texttt{gpt-4.1};
    Section~\ref{sec:judge-method}) is itself a model with its own biases; we mitigate
    obvious confounds (blinding to model identity, mode, and \texttt{dart\_pass}
    outcome) but do not independently validate judge scores against human raters in
    this work. A human-agreement spot-check on a sample of judged rows would
    strengthen this result and is left to future work.
    \item \textbf{Small-$n$ subgroup analyses.} The category-level win analysis
    (Section~\ref{sec:category}) is based on 63--86 total wins per architecture; the two
    highlighted categories are robust across both architectures, but per-category counts
    within that are small and we do not report category-level significance tests.
    \item \textbf{\texttt{qwen-direct} trained under an unintended, non-decaying
    learning rate.} As disclosed in Section~\ref{sec:trainingdata}, a scheduler
    configuration error left \texttt{qwen-direct}'s learning rate at
    $2.9$--$3.0\times10^{-5}$ for the back half of its run, instead of annealing down to
    the $3\times10^{-6}$ floor the other three arms reached. Consistent with this, its
    validation loss falls steadily through step $\sim$2{,}400 and then plateaus into a
    noisy $\approx0.143$--$0.164$ band for the remaining $\approx$2{,}600 steps rather
    than continuing to settle --- the kind of persistent oscillation an LR that never
    anneals would be expected to produce. We take this as suggestive, not conclusive,
    evidence that a properly-decaying run would have converged to a tighter optimum.
    Despite the deviation, \texttt{qwen-direct} is the best-performing arm in this paper
    on every metric we report (Table~\ref{tab:aggregate}), so we have no evidence it
    hurt the result; if a corrected schedule would have pushed performance higher still,
    that only strengthens our central finding rather than undermining it. We report the
    result from the checkpoint actually produced rather than re-running training, and
    flag this explicitly for reproducibility.
\end{itemize}

\section{Conclusion}
\label{sec:conclusion}

Across two architecturally distinct code models trained on the same Flutter/Dart editing
data, direct whole-file generation outperforms iterative diff-based generation on every
metric we measured, including a blinded LLM judge's assessment of code that compiles
successfully under both regimes. This is not primarily a story about the diff-based model
running out of budget or producing unparseable edits --- most of the gap comes from
trajectories that complete normally but drift semantically, disproportionately so when an
edit's \texttt{search} target is ambiguous. At the same time, diff-based generation is not
uniformly worse: it is specifically competitive on short, spatially local edits, and the
categories where it wins are exactly the categories that are, independently, shortest in
required trajectory length. We term this task locality and suggest it as the right axis
along which to decide, per edit rather than per model, whether a whole-file or diff-based
generation strategy is appropriate --- a conclusion that agrees with, and adds an
independent empirical data point to, recent adaptive-format proposals
\citep{diff-or-not}.

\section{Data and Code Availability}
\label{sec:availability}

The four evaluation datasets underlying every table and figure in this paper are
released on the Hugging Face Hub, one per architecture/mode arm, joined on a shared
\texttt{sample\_id} column that identifies the same underlying task across all four:

\begin{itemize}[leftmargin=*]
    \item \url{https://huggingface.co/datasets/bbidpa/Rainbow-Pony-100m-Flutter-direct-eval} (1,789 rows)
    \item \url{https://huggingface.co/datasets/bbidpa/Rainbow-Pony-100m-Flutter-steps-eval} (1,789 rows)
    \item \url{https://huggingface.co/datasets/bbidpa/Qwen2.5-Coder-0.5B-Flutter-direct-eval} (1,792 rows)
    \item \url{https://huggingface.co/datasets/bbidpa/Qwen2.5-Coder-0.5B-Flutter-steps-eval} (1,792 rows)
\end{itemize}

Each dataset includes the task instruction, initial file, reference final file, and
model output, together with:
\begin{small}
\begin{itemize}[leftmargin=*]
    \item automated metrics: \texttt{dart\_pass} and the underlying \texttt{dart
    analyze} error/warning/info counts, \texttt{bits\_per\_byte},
    \texttt{similarity\_ratio};
    \item steps-mode trajectory metadata: \texttt{stop\_reason}, \texttt{num\_steps},
    \texttt{num\_fallback\_steps},\\ \texttt{action\_list}, \texttt{steps\_log};
    \item blinded LLM-judge scores: \texttt{judge\_goal\_fulfillment},
    \texttt{judge\_correctness}, \texttt{judge\_code\_quality}, \texttt{judge\_notes}
    (judge model: \texttt{gpt-4.1}).
\end{itemize}
\end{small}
The four fine-tuned model checkpoints, plus the shared Rainbow-Pony pretrained-only
base checkpoint (Section~\ref{sec:models}), are released at:

\begin{itemize}[leftmargin=*]
    \item \url{https://huggingface.co/bbidpa/Rainbow-Pony-100m-Flutter-base} (pretrained-only checkpoint, shared by both Rainbow-Pony arms)
    \item \url{https://huggingface.co/bbidpa/Rainbow-Pony-100m-Flutter-direct}
    \item \url{https://huggingface.co/bbidpa/Rainbow-Pony-100m-Flutter-steps}
    \item \url{https://huggingface.co/bbidpa/Qwen2.5-Coder-0.5B-Flutter-direct}
    \item \url{https://huggingface.co/bbidpa/Qwen2.5-Coder-0.5B-Flutter-steps}
\end{itemize}

The two fine-tuning datasets underlying Section~\ref{sec:trainingdata} are released at:

\begin{itemize}[leftmargin=*]
    \item \url{https://huggingface.co/datasets/bbidpa/flutter-full-examples-v1} (14,600 rows; direct-mode fine-tuning source)
    \item \url{https://huggingface.co/datasets/bbidpa/flutter-diff-steps-v1} (100K--1M rows; steps-mode fine-tuning source, decomposed from the same source examples)
\end{itemize}

Licensing differs by artifact type: the four evaluation datasets and all five model
checkpoints are released under the MIT license; the two fine-tuning datasets
(\texttt{flutter-full-examples-v1} and \texttt{flutter-diff-steps-v1}) are released
under the Apache-2.0 license.

The training and evaluation harness code --- fine-tuning scripts, the step-decomposition
pipeline that produces \texttt{flutter-diff-steps-v1}, the Dart static-analysis evaluation
scaffold, and the blinded judge script (\texttt{judge\_outputs.py}) --- is released at
\url{https://github.com/bbidpa/rainbow-pony}.

% ---------- bibliography ----------

\end{document}